\documentclass[sigconf]{acmart}
\AtBeginDocument{%
  }

\setcopyright{acmlicensed}
\copyrightyear{2018}
\acmYear{2018}
\acmDOI{XXXXXXX.XXXXXXX}
\acmConference[Conference acronym 'XX]{Make sure to enter the correct
  conference title from your rights confirmation email}{June 03--05,
  2018}{Woodstock, NY}
\acmISBN{978-1-4503-XXXX-X/2018/06}

\usepackage{mathrsfs}
\usepackage{amsmath}
\usepackage{pifont}
\usepackage{booktabs}
\usepackage{makecell}
\usepackage{array}
\usepackage{colortbl}

\begin{document}

\title{Large Reasoning Embedding Models: \\ Towards Next-Generation Dense Retrieval Paradigm}


\author{Jianting Tang}
\affiliation{
    \institution{Researcher}
    \city{Hefei}
    \state{Anhui}
    \country{China}
}
\email{jiantingtang@mail.ustc.edu.cn}

\author{Dongshuai Li}
\affiliation{%
  \institution{Taobao \& Tmall Group of Alibaba}
  \city{Zhejiang}
  \state{Hangzhou}
  \country{China}
}
\email{lidongshuai.lds@taobao.com}

\author{Tao Wen}
\affiliation{%
  \institution{Taobao \& Tmall Group of Alibaba}
  \city{Zhejiang}
  \state{Hangzhou}
  \country{China}
}
\email{wentao.wen@alibaba-inc.com}

\author{Fuyu Lv}
\affiliation{%
  \institution{Taobao \& Tmall Group of Alibaba}
  \city{Zhejiang}
  \state{Hangzhou}
  \country{China}
}
\email{fuyu.lfy@taobao.com}

\author{Dan Ou}
\affiliation{%
  \institution{Taobao \& Tmall Group of Alibaba}
  \city{Zhejiang}
  \state{Hangzhou}
  \country{China}
}
\email{oudan.od@taobao.com}

\author{Linli Xu}
\affiliation{
    \institution{Researcher}
    \city{Hefei}
    \state{Anhui}
    \country{China}
}
\email{linlixu@ustc.edu.cn}

\newcommand{\modelname}{LREM}
\newcommand{\red}[1]{\textcolor{red}{#1}}
\newcommand{\blue}[1]{\textcolor{blue}{#1}}


\begin{abstract}
In modern e-commerce search systems, dense retrieval has become an indispensable component. By computing similarities between query and item (product) embeddings, it efficiently selects candidate products from large-scale repositories.
With the breakthroughs in large language models (LLMs), mainstream embedding models have gradually shifted from BERT to LLMs for more accurate text modeling.
However, these models still adopt direct‑embedding methods, and the semantic accuracy of embeddings remains inadequate.
Therefore, contrastive learning is heavily employed to achieve tight semantic alignment between positive pairs.
Consequently, such models tend to capture statistical co-occurrence patterns in the training data, biasing them toward shallow lexical and semantic matches.
For difficult queries exhibiting notable lexical disparity from target items, the performance degrades significantly.
In this work, we propose the \textbf{L}arge \textbf{R}easoning \textbf{E}mbedding \textbf{M}odel (\textbf{\modelname{}}), which novelly integrates reasoning processes into representation learning. For difficult queries, \modelname{} first conducts reasoning to achieve a deep understanding of the original query, and then produces a reasoning-augmented query embedding for retrieval. This reasoning process effectively bridges the semantic gap between original queries and target items, significantly improving retrieval accuracy.
Specifically, we adopt a two-stage training process: the first stage optimizes the LLM on carefully curated Query-CoT-Item triplets with SFT and InfoNCE losses to establish preliminary reasoning and embedding capabilities, and the second stage further refines the reasoning trajectories via reinforcement learning (RL).
Extensive offline and online experiments validate the effectiveness of \modelname{}, leading to its deployment on China's largest e-commerce platform since August 2025.

\end{abstract}



\received{20 February 2007}
\received[revised]{12 March 2009}
\received[accepted]{5 June 2009}

\maketitle

\section{Introduction}


With the advancement of deep learning, dense retrieval~\citep{dr1,dr2,dr3} has become an indispensable component of modern e-commerce search systems. By leveraging text embedding models pre‑trained on large‑scale corpora, dense retrieval is effective at capturing semantic relationships between query texts and item texts. During retrieval, it compares the query embedding against all pre-stored item embeddings to return the most relevant candidates.

The semantic accuracy of embeddings is critical to dense retrieval performance. In the past, BERT~\citep{bert}, RoBERTa~\citep{roberta}, and T5~\citep{t5} were widely adopted as text embedding models. 
Recently, LLMs such as LLaMA4~\citep{llama4}, Gemma3~\citep{gemma3}, and Qwen3~\citep{qwen3} have achieved significant breakthroughs in text understanding and generation.
Motivated by these advances, recent work has explored adapting LLMs into embedding models to leverage their extensive world knowledge and powerful language modeling capabilities.
For example, RepLLaMA~\citep{repllama} directly uses the last-token hidden state of LLaMA2~\citep{llama2} as the text embedding.
NV-Embed~\citep{nvembed} modifies the LLM architecture by introducing bidirectional attention and a novel latent attention layer to produce more accurate embeddings.

\begin{figure}[t!]
  \centering
  \includegraphics[width=\linewidth]{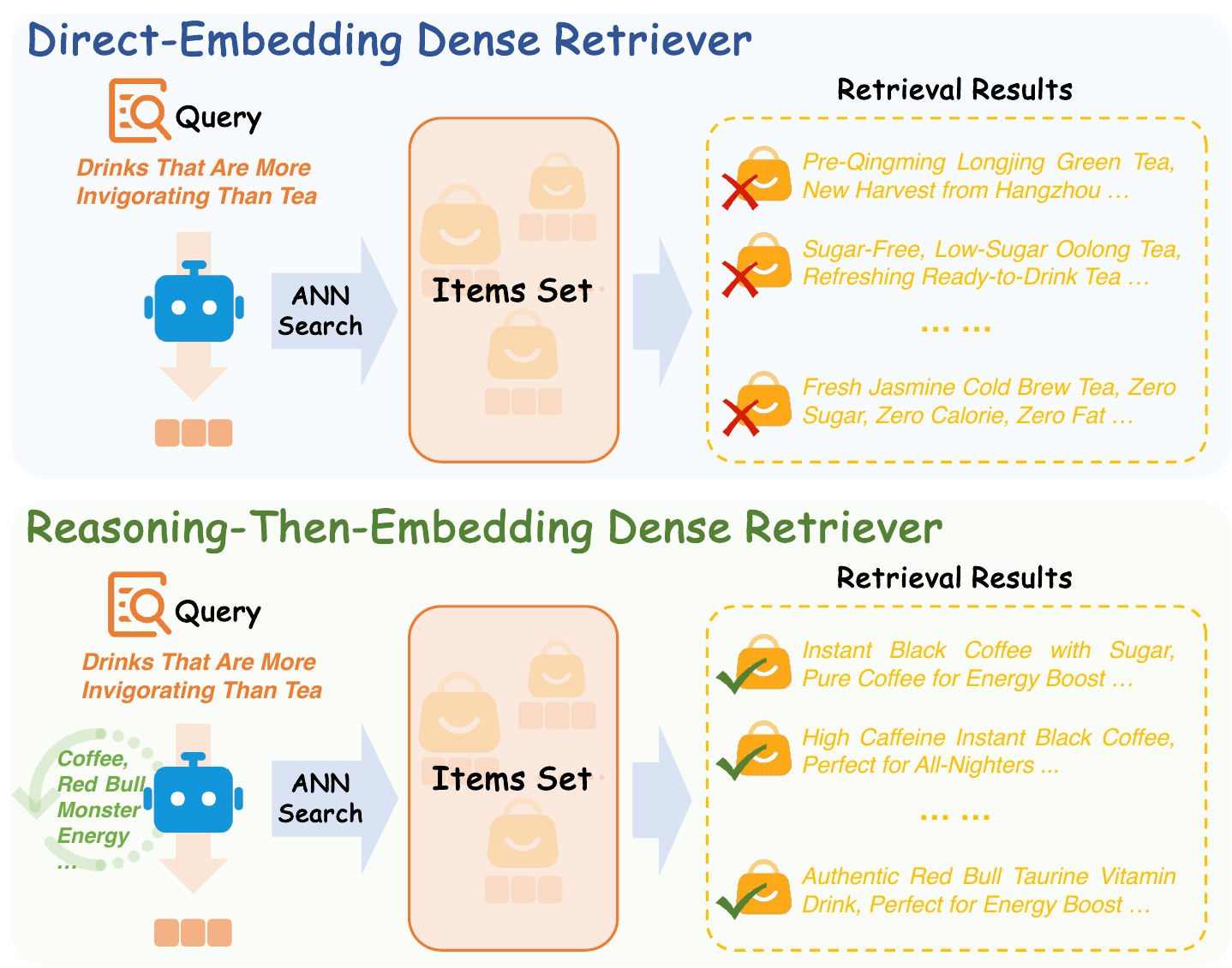}
  \vspace{-10pt}
  \caption{
   Comparison between traditional direct-embedding and the proposed reasoning-then-embedding dense retriever (\modelname{}). \modelname{} leverages reasoning to enable deep query understanding and accurate embeddings, overcoming the superficiality of direct-embedding methods.
  }
  \vspace{-10pt}
  \label{fig:page_begin}
\end{figure}


Despite leveraging more powerful LLMs, current dense retrievers still adopt direct-embedding methods~\citep{repllama,llama2vec,nvembed} that generate embeddings in a single forward pass, in which the semantic accuracy of embeddings remains significantly constrained.
To compensate, these models rely heavily on contrastive learning~\citep{cl} to forcibly align human-annotated positive query-item pairs in the embedding space, while pushing negative pairs apart.
Consequently, this paradigm encourages the model to exploit superficial co‑occurrence patterns in the training data and to engage in shallow lexical or semantic matching~\citep{bright,reasonir,rader}.
When handling difficult queries that exhibit notable lexical disparity from the target items, performance degrades significantly.
As illustrated in Figure~\ref{fig:page_begin}, for the query ``Drinks That Are More Invigorating Than Tea'', the direct-embedding dense retriever produces an inaccurate embedding, returning a large proportion of tea-based drinks rather than intended targets like Coffee or Red Bull.

As inherently generative models, LLMs excel at achieving precise semantic understanding of texts through explicit chain-of-thought (CoT) reasoning~\citep{cot,tot,got,deepseekmath}. However, harnessing this distinctive capability for dense retrieval remains largely underexplored in the current literature.
Therefore, to unlock the full potential of LLMs, we propose the Large Reasoning Embedding Model (\modelname{}), an entirely new paradigm that integrates reasoning into representation learning.
For difficult queries, where directly generating the embedding in a single forward pass results in poor semantic accuracy, \modelname{} performs an explicit reasoning process through CoT generation.
Consequently, the CoT serves as a semantic bridge, effectively linking original difficult queries with target items and ultimately enhancing retrieval performance. This reasoning-then-embedding paradigm advances the development of a more intelligent generation of dense retrievers.

To equip \modelname{} with both powerful reasoning and embedding capabilities, we develop a sophisticated data construction pipeline, as illustrated in Figure~\ref{fig:framework}. 
First, we collect queries from online logs—specifically those exhibiting poor performance under traditional direct-embedding dense retrievers. Then, we employ the advanced Qwen3-30B-A3B-Instruct~\citep{qwen3} LLM to generate a CoT for each query. To maximize information density and emphasize critical contents, each CoT is finally structured as a compact list of keywords. 
Then, we feed the query with CoT into a traditional dense retriever to obtain candidate items and filter the results with a relevance model. 
Consequently, this pipeline yields approximately 75.06 million Query-CoT-Item triples.
We adopt a two-stage training process. In the first stage, \modelname{} is jointly optimized via supervised fine-tuning (SFT) and InfoNCE losses, thereby acquiring preliminary reasoning and embedding capabilities. In the second stage, to further refine the reasoning trajectories and enhance CoT quality, we apply a GRPO-based~\citep{deepseekmath} RL algorithm while retaining the InfoNCE loss to maintain embedding alignment.

In summary, the contributions of this work are as follows:
\begin{itemize}

\item We propose \modelname{}, a next-generation dense retriever based on a novel reasoning-then-embedding paradigm, effectively overcoming the shallow semantic matching limitations of direct-embedding approaches.

\item We introduce an effective data construction pipeline and a two-stage training process that fully exploit the LLM's reasoning and embedding capabilities.

\item Extensive offline and online experiments demonstrate the effectiveness of our proposed \modelname{}, establishing a solid foundation for future research in dense retrieval.

\end{itemize}

\section{Related Work}
\subsection{Dense Retrieval}
In dense retrieval~\citep{dr4,dr5,dr6}, queries and items are encoded into a shared semantic space, and relevant items are returned based on embedding similarity using approximate nearest neighbor (ANN)~\citep{ann1,ann2,ann3} search algorithms.
Traditional dense retrieval methods primarily rely on fine-tuning pre-trained encoders such as BERT~\citep{bert}, RoBERTa~\citep{roberta}, or T5~\citep{t5} via contrastive learning to align text embeddings.
Recently, increasing studies have adopted LLMs as backbones~\citep{geminiemb,multilinemb,sfremb,gteemb} for dense retrieval, leveraging their superior semantic understanding capability and extensive world knowledge.
Among them, RepLLaMA~\citep{repllama} is the first to demonstrate the effectiveness of directly fine-tuning an open-source LLM for dense retrieval, establishing a strong baseline. 
Llama2Vec~\citep{llama2vec} applies unsupervised post-pretraining with novel embedding-oriented objectives (EBAE and EBAR) to further improve embedding quality. 
NV-Embed~\citep{nvembed} reconfigures the decoder-only architecture with bidirectional attention and an additional latent attention layer, enhancing embedding discriminability.
ICL-Embedder~\citep{iclembedder} further enhances generalization by explicitly training the model to leverage in-context examples during embedding generation, resulting in robust few-shot retrieval capabilities.
Currently, LLM-based embedding models have achieved SOTA performance across various text retrieval benchmarks, including the MTEB~\citep{mteb}. 
However, these models still follow the direct‑embedding method and perform poorly on difficult queries that requiring reasoning.

\subsection{LLM Reasoning}
Early methods—like Chain-of-Thought (CoT)~\citep{cot}, Tree-of-Thought (ToT)~\citep{tot}, and Graph-of-Thought (GoT)~\citep{got}—relied on human-crafted prompting strategies to guide LLMs in producing reasoning steps or exploring branching trajectories~\citep{reflection,mathprompter,autocap,planning}. However, within these frameworks, LLMs are unable to learn from prior explorations and to develop inherent reasoning capabilities. This limitation has driven the development of RL–based methods~\citep{limr,lean,prime,cpo} for eliciting LLM reasoning.
Proximal Policy Optimization (PPO)~\citep{ppo1,ppo2,ppo3,rlhf} serves as the foundational critic-based baseline, leveraging a learned value model to provide token-level advantage estimates but incurring substantial computational overhead. 
In verifiable reasoning settings, where reliable sequence-level rewards are available, research has increasingly favored critic-free variants that dispense with the value model entirely. 
Group Relative Policy Optimization (GRPO)~\citep{deepseekmath} exemplifies this shift by replacing critic-based advantages with group-normalized returns across multiple rollouts of the same prompt, reducing variance and stabilizing updates. 
Building on GRPO, Dynamic sAmpling Policy Optimization (DAPO)~\citep{dapo} incorporates dynamic sampling and ``clip-higher'' objectives to mitigate entropy collapse and concentrate computation on medium-difficulty prompts, thereby enhancing exploration efficiency.
Currently, RL~\citep{drgrpo,gspo,gmpo,mimo,shorterbetter} has offered a promising approach to training LLMs for sophisticated reasoning.

\subsection{Reasoning-Intensive Retrieval}
Reasoning-intensive retrieval~\citep{bright,diver,judgerank,frustra,elicit,reasoning} targets documents that cannot be found through simple lexical matching and require reasoning to bridge the semantic gap.
Some studies adopt a data-centric approach, directly constructing datasets of reasoning query–item pairs to train embedding models. RaDeR~\citep{rader} collects LLM reasoning trajectories in mathematical problem-solving, employing self-reflective relevance evaluation to generate diverse queries and challenging hard negatives, while ReasonIR~\citep{reasonir} synthesizes varied-length reasoning queries and negatives from seed documents through ReasonIR-Synthesizer, leading to SOTA performance on the BRIGHT~\citep{bright} benchmark.
Others focus on training specialized query rewriting models to refine queries prior to retrieval.
TongSearch-QR~\citep{tongsearch} trains small-scale query reasoning models via GRPO with a semi-rule-based reward,
while DeepRetrieval~\citep{deepretrieval} optimizes query rewriting for retrieval via PPO using retrieval metrics as rewards. 
Some work integrates reasoning and retrieval in an iterative manner~\citep{onegen,synergizing,grit}. R3‑RAG~\citep{r3rag} trains LLMs with both process and outcome rewards to adaptively perform multi-step reasoning and document retrieval, outperforming fixed human-designed workflows.
Compared with these complex query preprocessing and multi‑stage pipelines, we propose \modelname{}, an entirely new dense retriever that seamlessly integrates reasoning and embedding into a unified process.

\section{Methodology}
In this section, we first formalize the operational process of traditional direct-embedding dense retrievers and our proposed \modelname{} (§3.1). We then describe the construction pipeline for the training data (§3.2), followed by the presentation of \modelname{}'s two‑stage training process (§3.3 and §3.4).

\subsection{Preliminary}
Let $q_i$ denote a query and $\mathcal{D}= \{d_1, \dots, d_n \}$ denote the collection of candidate items. A dense retriever $f_{\theta}$ aims to project both the query and all candidate items into a shared semantic embedding space, where the set of relevant items for $q_i$, represented as $D^{+}_{q_i} = \{ d^{+}_{q_i,1}, \dots, d^{+}_{q_i,m} \}$ with $m \ll n$, can be retrieved according to a similarity measure $s(\boldsymbol{q}_i, \boldsymbol{d}_j) \in \mathbb{R}$.

For traditional dense retrievers, the query embedding is directly computed as $\boldsymbol{q}_i = f_{\theta}(q_i)$.
In contrast, \modelname{} performs reasoning to deeply understand the query before deriving the embedding,
\begin{gather}
    c_{i} = f^{\mathrm{gen}}_{\theta}(q_i) = (t_1, t_2, \dots, t_{l_i}), \quad t_k \in \mathcal{V}, \\
    \boldsymbol{q}_i = f^{\mathrm{emb}}_{\theta}([q_i; c_{i}]),
\end{gather}
where $l_i$ is the number of tokens in the generated CoT $c_{i}$, and $\mathcal{V}$ denotes the vocabulary of the LLM. The $c_{i}$ is concatenated with the original $q_i$ to form the reasoning-augmented query $[q_i; c_{i}]$ and then encoded into the final query embedding $\boldsymbol{q}_i$.

With item embeddings pre-computed directly as $\boldsymbol{d}_j = f_{\theta}(d_j)$, retrieval proceeds by ranking candidates in $\mathcal{D}$ according to the similarity measure and selecting the top-K results,
\begin{align}
\mathcal{D}_{q_i} = \operatorname{TopK}_{d_j \in \mathcal{D}} s(\boldsymbol{q}_i, \boldsymbol{d}_j).
\end{align}

\begin{figure*}[t!]
  \centering
  \includegraphics[width=\linewidth]{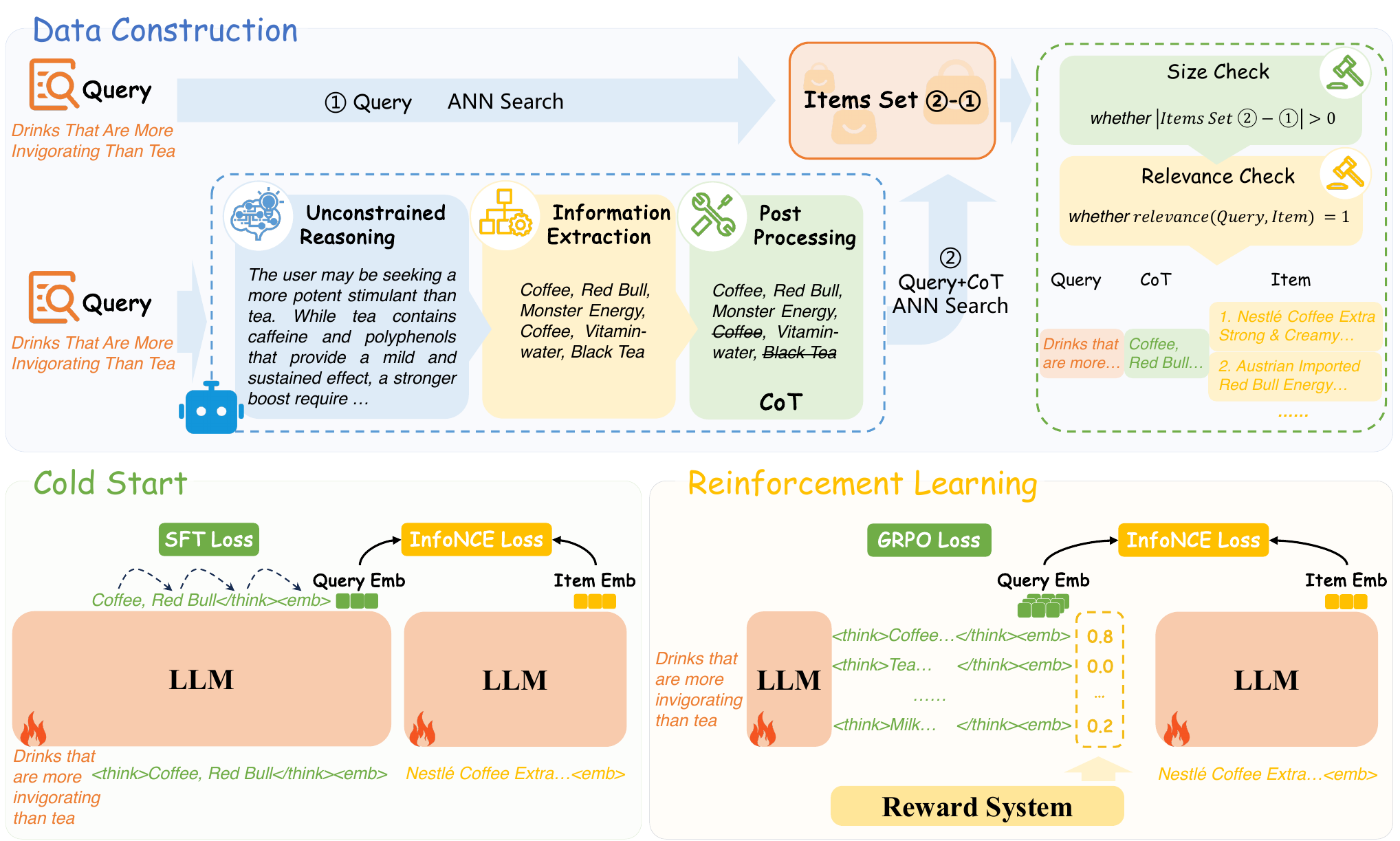}
  \vspace{-10pt}
  \caption{
    (1) Data Construction: An LLM generates keyword-based CoTs for each query. By comparing the retrieval results of a traditional dense retriever with and without the CoT, we discard queries where the CoT provides no gains, and filter truly relevant items via an advanced relevance model for the remaining queries. (2) Cold Start: \modelname{} is trained on Query-CoT-Item triplets, where the SFT loss optimizes \modelname{}'s reasoning process and the InfoNCE loss aligns the reasoning‑augmented query embedding with the item embedding. (3) Reinforcement Learning: \modelname{} is trained on Query-Item pairs, where the GRPO loss encourages exploration of superior reasoning trajectories under the guidance of a reward system, and the InfoNCE loss concurrently aligns the embeddings. Note that the same \modelname{} is applied to both the query-side and the item-side embedding.}
  \vspace{-5pt}
  \label{fig:framework}
\end{figure*}

\subsection{Data Construction}
Given an input query, \modelname{} first engages in reasoning to ensure a deep understanding of its semantics before generating the corresponding query embedding, thereby mitigating inaccuracies arising from the direct-embedding approach. Therefore, \modelname{} needs to possess both reasoning and embedding capabilities.
We adopt a two-stage training paradigm: Stage 1 uses carefully constructed Query-CoT-Item triplets for a cold-start training to establish preliminary reasoning and embedding capabilities; Stage 2 applies reinforcement learning to further optimize the reasoning trajectory.
This section elaborates on the construction pipeline for the Query-CoT-Item triplet dataset.
\subsubsection{CoT Generation}
We employ a large-parameter 30B Mixture-of-Experts (MoE) LLM~\citep{qwen3} to generate high-quality CoT for queries collected from online logs, thereby enabling this powerful model to teach \modelname{} how to reason. 
Since \modelname{} generates CoT sequentially during online retrieval, general CoTs—formatted as fluent, natural language sentences—can be verbose and introduce significant latency, potentially leading to request timeouts. 
To mitigate this, we structure each CoT training sample as a compact list of keywords.
Figure~\ref{fig:framework} illustrates the dedicated prompt framework for the CoT construction process, with more details provided in the Appendix. 
First, we prompt the LLM to perform unconstrained reasoning on the given query, without imposing any restrictions on output format or style, in order to preserve the model’s maximum reasoning capacity.
We then feed both the original query and the unconstrained reasoning output back into the LLM, prompting it to perform information extraction to identify relevant keywords.
Finally, we apply rule-based post-processing to the extracted results to remove duplicate keywords, keywords overlapping with the query, and any prohibited keywords.
To this point, we derive a compact CoT for each query that effectively captures its semantics.

\subsubsection{Item Filtering}
For difficult queries, such as ``Drinks That Are More Invigorating Than Tea'', collecting correctly matched items to form the training data is non-trivial.
If a traditional dense retriever directly encodes the original query into an embedding and retrieves the top‑$k$ items based on embedding similarity, most of the returned items are irrelevant. We denote this retrieved results as set \ding{172}.
Therefore, we combine the original query with the previously constructed CoT to jointly obtain the query embedding through the same retriever, and denote the retrieved top‑$k$ items as set \ding{173}.
By leveraging additional CoT information, set \ding{173} contains a substantially higher proportion of relevant items.
Subsequently, we further process the two sets.
We obtain the difference set \ding{173}-\ding{172}, in which all items are retrieved due to the incorporation of CoT information, indicating that the CoT is closely related to the textual content of these items.
First, we examine the size of the difference set. If it is 0, indicating that incorporating CoT for the query has no effect on retrieval results and yields no performance gain, this query is directly discarded.
Otherwise, we employ our internal advanced relevance model TaoSR1~\citep{taosr1}—a 42B MoE LLM trained via multi-stage RL and equipped with thinking capabilities—to assess the relevance between the original query and each item in set \ding{173}-\ding{172}, retaining only items that are judged as relevant.
Therefore, in the final constructed Query-CoT-Item triplets, the CoT plays a pivotal role in accurately retrieving the target items.
Such data are randomly split into two subsets: one for the cold-start stage using full Query-CoT-Item triplets, and one for the RL stage using only Query-Item pairs.

\subsection{Cold Start}
To enable \modelname{} to reason in the desired keyword-based format and to generate effective embeddings for dense retrieval, we conduct cold-start training on the constructed Query-CoT-Item triplets.
Meanwhile, each CoT sample is truncated to a maximum length of $l$, training \modelname{} to efficiently reason over key information within the length limit and avoid request timeouts during online retrieval.
Specifically, we introduce three special tokens—<think>, </think>, and <emb>—into \modelname{}'s vocabulary, and employ SFT loss to train the model to reason in the format ``<think> Specific CoT </think><emb>''. As the CoTs in the training data are produced by an advanced 30B MoE LLM, this process effectively distills the teacher model's superior reasoning capabilities into \modelname{}.
Formally, the training loss for this process is defined as a standard next‑token prediction objective:
\begin{equation}
    \mathcal{L}_{\text{SFT}} = -\frac{1}{N} \sum_{i=1}^{N} \sum_{j=1}^{l_i} \log P(t_j \mid q_i, t_{<j}),
\end{equation}
where $q_i$ denotes the input query, $t_j$ is the $j$-th token in the CoT, $l_i$ is the total number of tokens in the CoT, and $N$ is the batch size.

Concurrently with the generative objective, we incorporate the in‑batch contrastive learning via the InfoNCE loss to minimize the embedding distance between positive query-item pairs. Specifically, the last hidden state of the <emb> token on the query side is adopted as the overall query embedding. Due to the LLM's causal attention mechanism, this embedding holistically integrates information from both the original query and the corresponding CoT, thereby possessing more accurate semantics. On the item side, we manually append an <emb> token to the product text title, and likewise use its last hidden state as the item embedding. Both the query and item embeddings are derived from the same \modelname{}.

Considering a mini-batch of $N$ training triples $\{(q_i, c_i, d_i)\}_{i=1}^N$, the process for obtaining the query embedding $\boldsymbol{q}_i$, the process for obtaining the item embedding $\boldsymbol{d}_i$, and the corresponding InfoNCE loss can be formally expressed as:
\begin{gather}
\boldsymbol{q}_i = \text{H}_\text{<emb>}^\text{[-1]} [\text{\modelname{}}( q_i \text{<think>} c_i \text{</think>}\text{<emb>} )], \\
\boldsymbol{d}_i = \text{H}_\text{<emb>}^\text{[-1]} [\text{\modelname{}}( d_i\text{<emb>} )], \\
\mathcal{L}_{\text{InfoNCE}} = -\frac{1}{N} \sum_{i=1}^N \log \frac{\exp( s(\boldsymbol{q}_i, \boldsymbol{d}_i) /\tau )}{\sum_{j=1}^N \exp( s(\boldsymbol{q}_i, \boldsymbol{d}_j) /\tau )},
\end{gather}
where $\text{\modelname{}}(\cdot)$ denotes the \modelname{} processing the input text and outputting a sequence of hidden states, and $\text{H}_\text{<emb>}^\text{[-1]}$ denotes selecting the last-layer hidden state of the <emb> token. $s(\boldsymbol{q}_i, \boldsymbol{d}_j) = \frac{\boldsymbol{q}_i^\top \boldsymbol{d}_j}{\|\boldsymbol{q}_i\| \|\boldsymbol{d}_j\|}$ computes the cosine similarity between $\boldsymbol{q}_i$ and $\boldsymbol{d}_j$, and $\tau$ denotes the temperature coefficient.

Finally, the total loss in the cold‑start stage is defined as:
\begin{equation}
    \mathcal{L} = \lambda_{1} \mathcal{L}_{\text{SFT}} + \lambda_{2} \mathcal{L}_{\text{InfoNCE}},
\end{equation}
where $\lambda_{1}$ and $\lambda_{2}$ are the loss coefficients.

\subsection{Reinforcement Learning}
After the cold-start stage, \modelname{} acquires preliminary reasoning and embedding capabilities. However, its reasoning ability is significantly constrained by the quality of the constructed CoT data. As the model is primarily engaged in imitation learning, this substantially hinders the full activation of its intrinsic reasoning capacity. Therefore, at this stage, we employ GRPO to encourage \modelname{} to conduct extensive reasoning exploration, fostering the generation of superior reasoning trajectories.
For each $q_i$, \modelname{} samples a group of $G$ CoTs, denoted as $\{c_i^g\}_{g=1}^G$, which are then assessed by the reward system from three dimensions: format, length and retrieval accuracy.
\modelname{} is encouraged to increase the generation likelihood of CoTs with higher reward scores.
\subsubsection{Format Reward}
During online inference, the query embedding is only accessible when \modelname{} generates CoTs in the correct ``<think> Specific CoT </think><emb>'' format. We therefore incorporate a format reward to encourage adherence.

\begin{equation}
r_{\text{format}} = 
\begin{cases} 
1 & \text{if the CoT meets the format specification} \\
0 & \text{otherwise}
\end{cases}.
\end{equation}

\subsubsection{Length Reward}
To avoid request timeouts caused by auto-regressive generation of excessively long CoTs, we encourage the model to limit its CoT length to at most $l$.

\begin{equation}
r_{\text{length}} = 
\begin{cases} 
1 & \text{If the CoT length is } \le l \\
0 & \text{otherwise}
\end{cases}.
\end{equation}


\subsubsection{Retrieval Accuracy Reward}
The ultimate objective of this task is to accurately retrieve the target item. Therefore, we propose a retrieval accuracy reward 
to guide \modelname{}'s exploration.
Specifically, for $q_i$ and associated $c_i^g$, we first obtain the corresponding query embedding $\boldsymbol{q}_i^g$. We then compute its cosine similarity with all item embeddings  $\{\boldsymbol{d}_j\}_{j=1}^N$ in the current batch, and rank these items by the similarity. 
The retrieval accuracy reward is based on the rank position of $d_i$ (i.e., the ground‑truth item paired with $q_i$). 
Higher ranks result in larger rewards, whereas lower ranks result in smaller ones. Formally, the reward is defined as,
\begin{gather}
\text{rank}(d_i) = 1 + \sum_{\substack{j=1, j \neq i}}^{N} \mathbb{I}\left(s(\boldsymbol{q}_i^g, \boldsymbol{d}_j) > s(\boldsymbol{q}_i^g, \boldsymbol{d}_i)\right), \\
r_{\text{accuracy}} = 1 - \frac{\log \text{rank}(d_i)}{\log N},
\end{gather}
where $\mathbb{I}$ is the indicator function, returning 1 if the condition holds and 0 otherwise.

Finally, the overall reward is defined as,
\begin{equation}
r = \beta_{1}r_{\text{format}} + \beta_{2}r_{\text{length}} + \beta_{3}r_{\text{accuracy}}
\end{equation}
where $\beta_{1}$, $\beta_{2}$ and $\beta_{3}$ are the reward coefficients.

\begin{table*}[t!]
    \centering
    \caption{Comparison of offline evaluation results across four challenging query categories. Overall results on all queries are in gray columns. The best results across all models are in bold, and the best baseline results are underlined. }
    \begin{tabular}{l|ccccc|ccccc}
        \toprule
        \textbf{Methods} & \multicolumn{5}{c|}{\textbf{HitRate@6000}} & \multicolumn{5}{c}{\textbf{Precision@100}} \\
        & {\small Q\&A} & {\small Alternative} & {\small Negative} & {\small Knowledge} & {\small \textbf{Overall}} & {\small Q\&A} & {\small Alternative} & {\small Negative} & {\small Knowledge} & {\small \textbf{Overall}} \\
        \midrule
        BERT & 11.73 & 30.38 & 34.40 & 23.30 & \cellcolor{gray!30}{24.96} & 69.60 & 26.86 & 57.40 & 50.49 & \cellcolor{gray!30}{51.09} \\
        \makecell[l]{Query-Rewrite} & 14.70 & 42.02 & 24.81 & 31.42 & \cellcolor{gray!30}{28.24} & 84.52 & 36.39 & 49.90 & 62.65 & \cellcolor{gray!30}{58.37} \\
        \makecell[l]{Qwen2.5 (Uni-Attn. Last)} & 14.61 & 42.20 & 39.54 & 36.24 & \cellcolor{gray!30}{32.52} & 86.20 & 36.17 & 64.05 & 68.44 & \cellcolor{gray!30}{65.38} \\
        \makecell[l]{Qwen2.5 (Uni-Attn. Mean)} & 14.47 & 42.05 & 39.33 & 36.18 & \cellcolor{gray!30}{32.38} & 86.14 & 36.02 & 63.90 & 68.23 & \cellcolor{gray!30}{65.24} \\
        \makecell[l]{Qwen2.5 (Uni-Attn. Latent)} & 14.76 & 42.43 & 39.65 & 36.44 & \cellcolor{gray!30}{32.69} & 86.06 & 35.91 & 63.84 & 68.09 & \cellcolor{gray!30}{65.14} \\
        \makecell[l]{Qwen2.5 (Uni-Attn. Ly4)} & 14.70 & 42.32 & 39.58 & 36.32 & \cellcolor{gray!30}{32.60} & 86.29 & 36.53 & 64.08 & 68.57 & \cellcolor{gray!30}{65.52} \\
        \makecell[l]{Qwen2.5 (Bi-Attn. Last)} & \underline{14.95} & \underline{42.54} & \underline{39.94} & \underline{36.61} & \cellcolor{gray!30}{\underline{32.89}} & \underline{86.35} & \underline{36.86} & \underline{64.16} & \underline{68.72} & \cellcolor{gray!30}{\underline{65.66}} \\
        \midrule
        \modelname{} (Cold Start) & 14.92 & 41.79 & 39.30 & 36.20 & \cellcolor{gray!30}{32.45} & 85.73 & 35.47 & 63.11 & 68.34 & \cellcolor{gray!30}{64.83} \\
        \modelname{} (Cold Start+RL) & \textbf{17.82} & \textbf{45.01} & \textbf{41.55} & \textbf{37.29} & \cellcolor{gray!30}{\textbf{34.78}} & \textbf{89.97} & \textbf{40.18} & \textbf{66.34} & \textbf{69.94} & \cellcolor{gray!30}{\textbf{68.22}} \\
        \bottomrule
    \end{tabular}
    \label{tab:offline}
\end{table*}

\subsubsection{Training Objective}
We adopt a GRPO‑based RL algorithm to optimize \modelname{}'s reasoning trajectory. The loss is defined as:
\begin{align}
\mathcal{L}_{\mathrm{GRPO}} 
&= -\mathbb{E} \bigg[
    \sum_{i=1}^G  
    \min\Bigg( 
        \frac{\pi_\theta(c_i|q)}{\pi_{\theta_\mathrm{old}}(c_i|q)}\, A_i, \notag\\
&\qquad\qquad
        \mathrm{clip}\!\left(
            \frac{\pi_\theta(c_i|q)}{\pi_{\theta_\mathrm{old}}(c_i|q)},\ 
            1-\epsilon,\ 
            1+\epsilon
        \right) A_i
    \Bigg)
\bigg],
\end{align}
where $A_i = \frac{r_i - \text{mean}(\{r_1, r_2, \cdots, r_G\})}{\text{std}(\{r_1, r_2, \cdots, r_G\})}$ denotes the advantage.

Concurrently, we employ the InfoNCE loss to dynamically align all reasoning-augmented query embeddings $\{\boldsymbol{q}_i^{g}\}_{g=1}^G$ corresponding to $q_i$, with the target item embedding $\boldsymbol{d}_i$. The loss is defined as:
\begin{equation}
\mathcal{L}_{\text{InfoNCE}} = - \frac{1}{NG} \sum_{i=1}^{N}\sum_{g=1}^{G} 
\log \frac{\exp \left( s(\boldsymbol{q}_i^{g}, \boldsymbol{d}_i) / \tau \right)}
{\sum_{j=1}^{N} \exp \left( s(\boldsymbol{q}_i^{g}, \boldsymbol{d}_j) / \tau \right)},
\end{equation}
where $G$ denotes the sampling group size.

Finally, the total loss in the RL stage is defined as:
\begin{equation}
    \mathcal{L} = \gamma_{1} \mathcal{L}_{\text{GRPO}} + \gamma_{2} \mathcal{L}_{\text{InfoNCE}},
\end{equation}
where $\gamma_{1}$ and $\gamma_{2}$ are the loss coefficients.

\section{Experiments}
\subsection{Experimental Setup}
\subsubsection{Dataset}
By using Qwen3-30B-A3B-Instruct~\citep{qwen3} to construct CoT for each query and an advanced relevance model TaoSR1~\citep{taosr1} to identify relevant items, we ultimately obtain 75.06 million Query-CoT-Item triples.
We randomly set aside 4 million Query-Item pairs for the RL stage, and use the remaining data for the cold‑start stage.
To rigorously assess \modelname{}'s performance, the test set predominantly includes four extremely challenging query categories: question-answering (Q\&A), affordable alternative, negative, and knowledge-intensive queries. In total, the test set contains 7209 queries, with a candidate product pool of 76.63 million items.
More details are provided in the Appendix.

\subsubsection{Metrics}
In offline experiments, the performance is evaluated based on Hit Rate~\citep{hr} and Precision~\citep{precision} metrics, and in online experiments, the GSB (Good/Same/Bad)~\citep{gsb} metric is adopted.
\begin{itemize}
\item HitRate@6000: It measures the ratio of ground‑truth items that appear within the top 6000 retrieved items, relative to the total number of ground‑truth items.
\item Precision@100: It evaluates the proportion of the top 100 retrieved items that are judged as relevant by TaoSR1.
\item GSB: It evaluates the superiority of a test bucket against the base bucket in A/B experiments by having human assessors compare the test model's retrieval results with the base model's for the same queries side-by-side. 
GSB $+x\%$ indicates that the test bucket outperforms the base bucket on $x\%$ of queries.
\end{itemize}

\subsubsection{Implementation Details}
We adopt Qwen2.5‑3B‑Instruct~\citep{qwen25} and conduct training on 128 GPUs.
During the cold start stage, CoT training samples are truncated to a maximum length of $l=16$. The loss coefficients are set to $\lambda_1=0.1$ and $\lambda_2=1$, with a per-GPU batch size of 128, a learning rate of 1e-5, and a cosine scheduler with a warmup ratio of 0.03. The model is trained for 1 epoch with all parameters optimized.
In the reinforcement learning stage, $G=8$ CoTs are sampled for each query. In the length reward, the threshold is set to $l=16$, and reward coefficients are configured as $\beta_1=0.5$, $\beta_2=0.2$ and $\beta_3=1$. The loss coefficients are set to $\gamma_1=1$ and $\gamma_2=0.1$, with a per-GPU batch size of 256, a learning rate of 1e-6, and a cosine scheduler with a warmup ratio of 0.03. The model is trained for 1 epoch with all parameters updated.

\subsubsection{Baselines}
We adopt Qwen2.5‑3B‑Instruct as the base model and reproduce a range of mainstream methods and configurations to serve as baselines. All baselines are trained for one epoch on the full 75.06 million Query-Item pairs.
\begin{itemize}
\item BERT~\citep{retromae}: A 12-layer 110M BERT trained via the RetroMAE~\citep{retromae} method and contrastive learning.
\item Query‑Rewrite~\citep{csaqr}: Query rewriting via CSA-QR~\citep{csaqr}, followed by retrieval via inverted index.
\item Qwen2.5 (Uni-Attn. Last)~\citep{repllama}: Unidirect-attention; embedding from last token's final-layer hidden state.
\item Qwen2.5 (Uni-Attn. Mean)~\citep{geminiemb}: Unidirect-attention; embedding from mean of all final-layer token hidden states.
\item Qwen2.5 (Uni-Attn. Latent)~\citep{nvembed}: Unidirect-attention; embedding from NV-Embed's latent attention layer.
\item Qwen2.5 (Uni-Attn. Ly4)~\citep{bertred}: Unidirect-attention; embedding from mean of last token's final 4 layer hidden states.
\item Qwen2.5 (Bi-Attn. Last)~\citep{gteemb}: Bidirect-attention; embedding from last token's final-layer hidden state.
\end{itemize}

\begin{figure*}[t!]
  \centering
  \includegraphics[width=\linewidth]{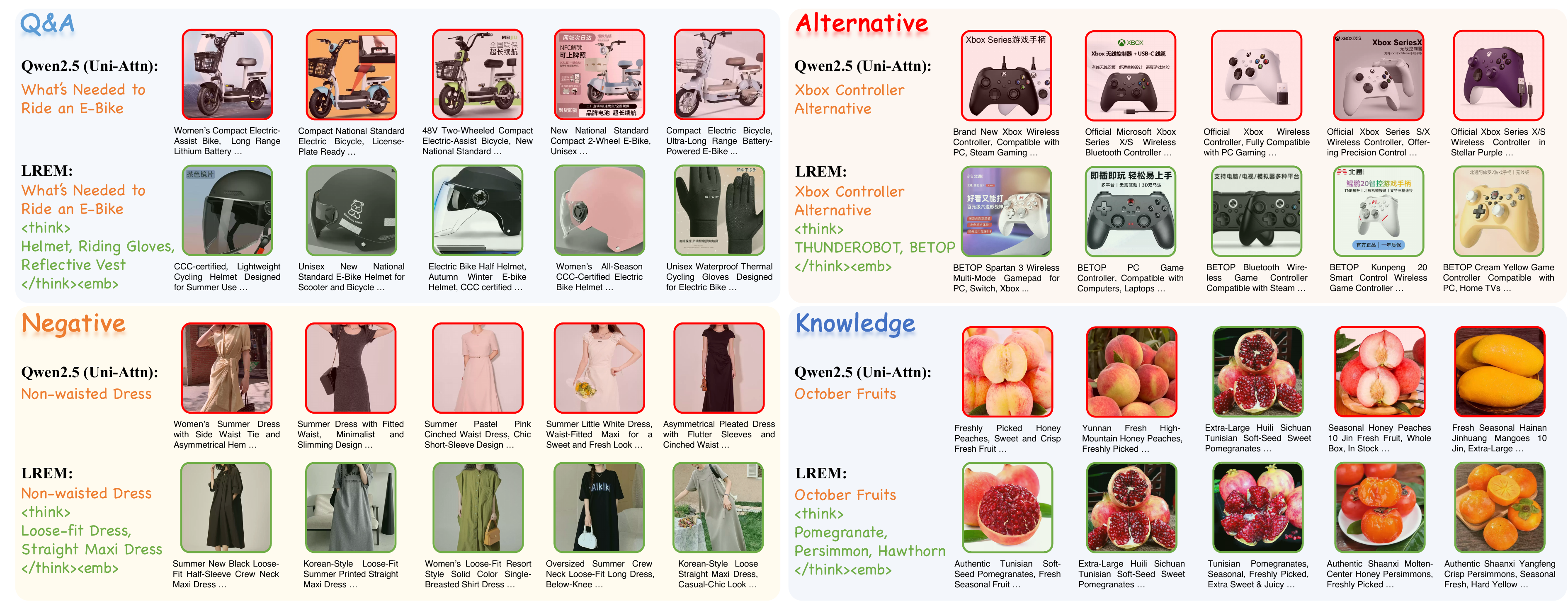}
  \caption{
    Representative examples from the four query categories. Compared with the direct-embedding method Qwen2.5 (Uni-Attn. Last), which performs only superficial lexical matching, \modelname{} adopts a novel reasoning-then-embedding paradigm, enabling accurate query semantic understanding and target item retrieval.
    }
  \label{fig:offlinecase}
\end{figure*}

\subsection{Offline Experiments}
\subsubsection{Main Results}
Offline evaluations are conducted based on the HitRate@6000 and Precision@100 metrics. 
As shown in Table~\ref{tab:offline}, \modelname{} trained with both cold-start and reinforcement learning stages achieves the best performance, outperforming the strongest baseline, Qwen2.5 (Bi-Attn. Last), by 5.75\% on HitRate@6000 and by 3.90\% on Precision@100 over all queries. 
Notably, the Q\&A and Alternative categories achieve the most substantial improvements, with gains of $19.20\%$ and $5.81\%$ on HitRate@6000, and $4.19\%$ and $9.01\%$ on Precision@100, respectively.
This improvement clearly highlights the superiority of \modelname{}'s reasoning-then-embedding paradigm.
Qwen2.5 (Uni-Attn. Last) outperforms BERT, primarily due to its larger parameter scale and extensive next‑token-prediction pre‑training on broad corpora. However, its unidirectional attention mechanism prevents earlier tokens from incorporating information from subsequent ones, which to some extent constrains the accuracy of semantic embeddings. Therefore, Qwen2.5 (Bi‑Attn. Last) achieves slightly better performance.
Since \modelname{} requires both reasoning and embedding capabilities, it retains a unidirectional attention architecture. Despite this constraint, it achieves the best performance, demonstrating that explicit reasoning effectively overcomes the fundamental flaw of superficial semantic understanding in direct-embedding methods. 
Compared with the Query‑Rewrite method, query rewriting inevitably incurs information loss from the original query, and the accumulation of errors across multiple stages further limits the final retrieval performance. In contrast, as a unified model, \modelname{} fosters deeper semantic understanding by incorporating reasoning before generating the final embedding, effectively bridging the semantic gap between original queries and target items.

\subsubsection{Case Studies}
As illustrated in Figure~\ref{fig:offlinecase}, we present several representative examples to illustrate the effectiveness of \modelname{}.
For queries ``What's Needed to Ride an E‑Bike'', ``Xbox Controller Alternative'', and ``Non‑waisted Dress'', the direct-embedding method Qwen2.5 (Uni‑Attn. Last) fails to capture the underlying intent of the original query and merely performs superficial lexical matching. Consequently, it incorrectly retrieves items whose textual titles contain keywords like ``E‑Bike'', ``Xbox Controller'', and ``Waisted Dress''.
In contrast, \modelname{} conducts explicit reasoning on the query before generating embeddings, thoroughly uncovering the underlying product‑search intent behind the queries. Specifically, it infers the first query likely requires items such as ``Helmet, Riding Gloves, Reflective Vest''; the second suggests suitable substitutes like ``THUNDEROBOT, BETOP''; and the third refers to ``Loose‑fit Dress, Straight Maxi Dress''. 
Through this reasoning process, \modelname{} ultimately generates accurate embeddings and successfully retrieves the correct items.
For the query ``October Fruits'', Qwen2.5 (Uni‑Attn. Last) naively retrieves various fruits, many of which are not typically harvested in October, indicating its limited understanding of temporal semantics. In contrast, \modelname{} accurately captures the seasonal context expressed in the original query.

\begin{table}[t!]
\centering
\caption{Retrieval performance under different manual modifications of \modelname{}'s CoT content.}
\vspace{-5pt}
\begin{tabular}{lccc}
    \toprule
    \textbf{Methods} & \textbf{HitRate@6000} & \textbf{Precision@100} \\
    \midrule
    \modelname{} & \textbf{34.78} & \textbf{68.22} \\
    \modelname{} (Empty-CoT) & 31.59 & 64.25 \\
    \modelname{} (Random-CoT) & 30.16 & 62.32 \\
    \modelname{} (Query-CoT) & 32.54 & 65.63 \\
    \bottomrule
\end{tabular}
\vspace{-8pt}
\label{tab:cotcontent}
\end{table}

\subsection{Ablation Experiments}
\subsubsection{Effect of Reinforcement Learning}
\begin{figure}[t!]
  \centering
  \includegraphics[width=\linewidth]{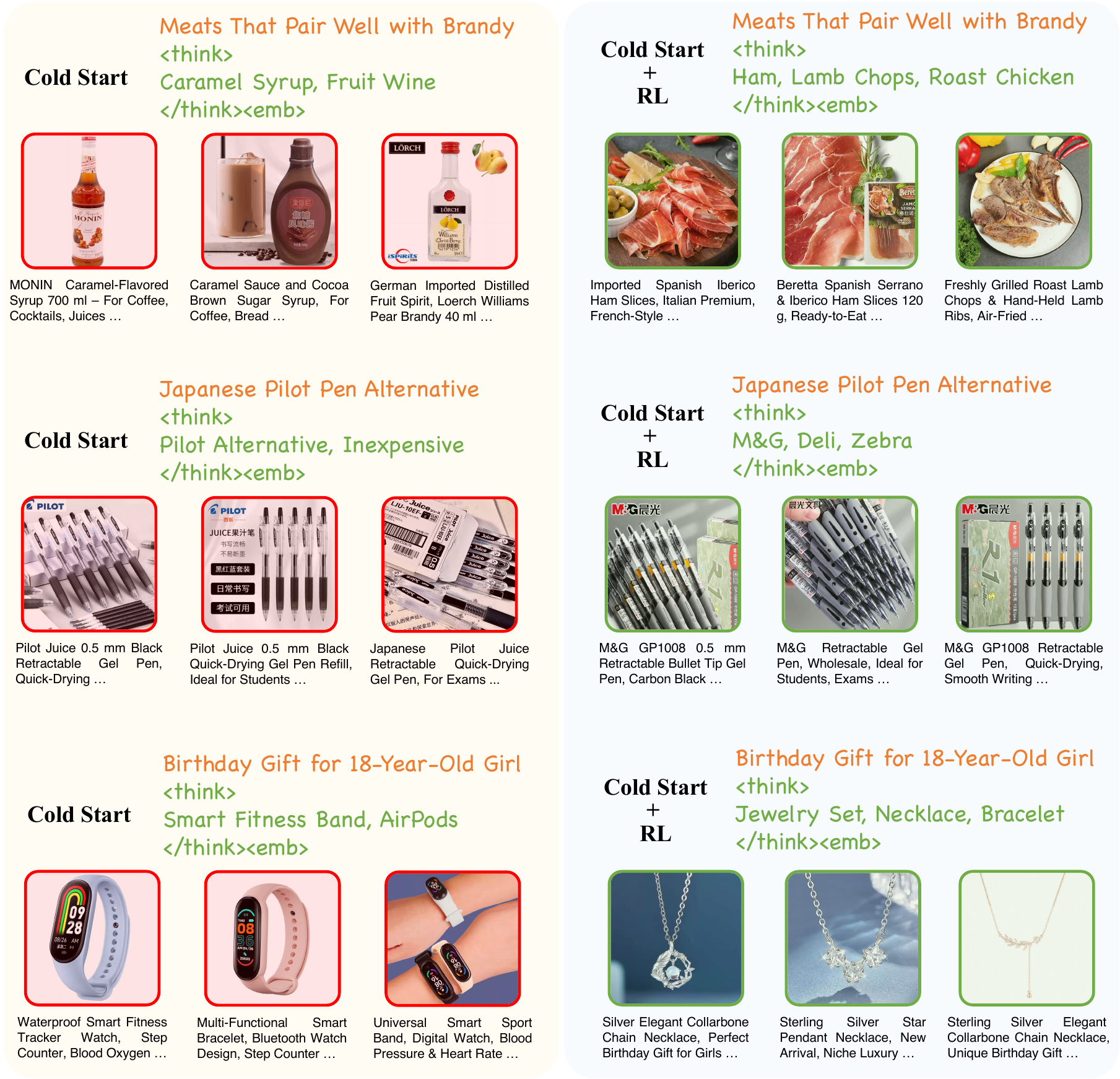}
  \vspace{-20pt}
  \caption{
    Comparison of generated CoT and retrieval results between \modelname{} (Cold Start) and \modelname{} (Cold Start + RL).}
  \vspace{-17.5pt}
  \label{fig:cotcompare}
\end{figure}

In the cold start stage, \modelname{} develops preliminary reasoning and embedding capacities. In this section, we assess the effect of RL. As shown in Table~\ref{tab:offline}, RL contributes critically to the final performance gains, substantially improves retrieval effectiveness across various types of queries, with an overall gain of 7.18\% in HitRate@6000 and 5.23\% in Precision@100.
Representative examples are presented in Figure~\ref{fig:cotcompare}.
For the query ``Meats That Pair Well with Brandy'', the cold-start model exhibits \textit{\textbf{incorrect reasoning}}, which misguides embedding generation and retrieves entirely irrelevant items. After RL, \modelname{} performs more accurate reasoning, enabling correct item retrieval.
For ``Japanese Pilot Pen Alternative'', the cold-start model performs \textit{\textbf{ineffective reasoning}}—``Pilot Alternative, Inexpensive''—which fails to delve into the original query and reason out additional valuable information, leading to persistent retrieval errors. In contrast, the RL-enhanced \modelname{} effectively follows the reasoning trajectory involving ``M\&G, Deli, Zebra'', returning intended items.
For ``Birthday Gift for 18-Year-Old Girl'', the cold-start model conducts \textit{\textbf{suboptimal reasoning}}, suggesting plausible but generic items such as ``Smart Fitness Band, AirPods''. RL enables \modelname{} to conduct more targeted reasoning—``Jewelry Set, Necklace, Bracelet''—better aligning with the intended gifting context.
In the cold‑start stage, \modelname{} primarily relies on imitation learning and is constrained by the quality of the constructed CoT data. RL effectively unlocks \modelname{}'s inherent capacity for deeper query reasoning and facilitates the exploration of superior reasoning trajectories.

\subsubsection{Effect of CoT Content}
The CoT generates by \modelname{} offers a transparent view into the model's internal process of query understanding.
To quantitatively assess the impact of CoT content on both embedding accuracy and retrieval performance, We deliberately prevent \modelname{} from generating the CoT on its own.
Instead, we manually construct three CoT variants: (1)\textit{Empty-CoT}, without any reasoning content (i.e., Query<think></think><emb>); (2) \textit{Random-CoT}, containing a sequence of random tokens (i.e., Query<think>Random Tokens</think><emb>); and (3) \textit{Query-CoT}, which simply repeats the original query within the reasoning field (i.e., Query<think>Query</think><emb>).
As shown in Table~\ref{tab:cotcontent}, \modelname{} (Empty-CoT) exhibits a substantial performance drop compared to \modelname{}—HR@6000 decreases by 9.17\% and Precision@100 decreases by 5.82\%—because it essentially degenerates into traditional direct‑embedding methods, preventing the model from developing a deep understanding of the query before generating its embedding.
\modelname{} (Random‑CoT) exhibits the worst performance, as it introduces substantial noise unrelated to the original query.
Although \modelname{} (Query-CoT) does not contribute to achieving a deeper level of semantic understanding, the simple repetition of the original query, to some extent, alleviates the limitations of unidirectional attention—where earlier tokens cannot attend to later ones—thereby yielding marginally better performance compared to \modelname{} (Empty-CoT), with +0.95 points in HR@6000 and +1.38 points in Precision@100.

\begin{figure}[t!]
  \centering
  \includegraphics[width=\linewidth]{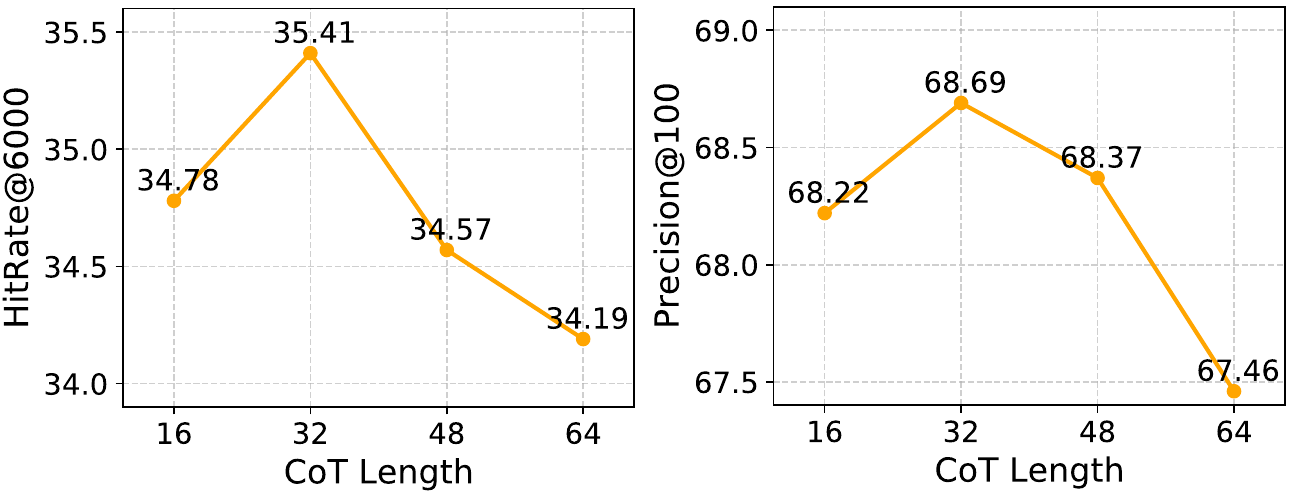}
  \caption{Retrieval performance across varying CoT lengths generated by \modelname{}.}
  \Description{Retrieval performance across varying CoT lengths.}
  \vspace{-5pt}
  \label{fig:cot_len}
\end{figure}

\subsubsection{Effect of CoT Length}
To investigate the influence of CoT length on retrieval performance, we train \modelname{} variants with different reasoning lengths by adjusting $l$.
As shown in Figure~\ref{fig:cot_len}, extending the CoT length in \modelname{} from 16 to 32 tokens yields a noticeable improvement in retrieval performance, as overly restrictive length constraints can hinder the model's reasoning process, leading to an incomplete understanding of the query.
However, further increasing the CoT length to 48 or 64 tokens leads to a decline in performance. This is attributed to the fact that, \modelname{} conducts reasoning in a keyword‑centric manner, excessively long and divergent keyword sequences may diminish semantic precision and distract the model's focus, thereby introducing difficulties for the final embedding. 
In our final models, $l$ is set to 16 to achieve a balance between retrieval performance and computational efficiency.

\subsection{Online Experiments}
\begin{table}[h!]
\centering
\vspace{-5pt}
\caption{Online A/B testing results.}
\label{tab:results}
\vspace{-10pt}
\begin{tabular}{lcccc}
    \toprule
    & Q\&A & Alternative & Negative & Knowledge \\
    \midrule
    \textbf{GSB} & +7.39\% & +7.27\% & +15.7\% & +4.94\% \\
    \bottomrule
\end{tabular}
\end{table}
We conduct an online A/B test via side-by-side human evaluations on 2000 live queries (four categories), comparing the experimental bucket (\modelname{}) against the base bucket (the current best online model). Results show consistent improvements: +7.39\% on Q\&A, +7.27\% on Alternative, +15.7\% on Negative, and +4.94\% on Knowledge queries.
Due to the additional reasoning process incorporated by \modelname{} during online retrieval, it incurs extra latency, with the average retrieval time rising from 15ms to 50ms relative to the base bucket.
Within the maximum allowable latency, \modelname{} trades longer processing time for more accurate retrieval.
In e‑commerce search scenarios, where queries often involve subtle preferences or implicit needs, \modelname{}'s reasoning‑then‑embedding approach effectively links user queries with intended products, leading to more precise and meaningful retrieval.

\section{Conclusion}
This paper proposes \modelname{}, a novel reasoning‑then‑embedding dense retriever. In contrast to traditional direct‑embedding models, \modelname{} first performs reasoning over the original query to ensure a deep semantic understanding, and then produces the corresponding embeddings. By introducing the explicit reasoning process, \modelname{} goes beyond superficial lexical matching and achieves superior retrieval performance on highly difficult queries—such as Q\&A, alternative, negative, and knowledge types. Extensive offline and online experiments demonstrate \modelname{}'s impressive performance, positioning \modelname{} as an important step toward the development of next‑generation, more intelligent embedding models.


\bibliographystyle{ACM-Reference-Format}
\balance
\bibliography{main}

%
\appendix

\begin{table*}[h!]
\centering
\small
\caption{Detailed prompts for instructing Qwen3‑30B‑A3B‑Instruct in unconstrained reasoning.}
\vspace{-10pt}
\begin{tabular}{p{0.9\linewidth}}
    \toprule
    \textbf{System Prompt} \\
    You are a query assistant for a Chinese e-commerce search platform, highly adept at understanding various user queries. \\
    \midrule
    \textbf{User Prompt} \\
    The following are examples illustrating the interpretation of several common query types: \\
    \\
    1. Substitute-seeking queries: For instance, when a user inputs ``ysl100 dupe'', the intent should be understood as seeking a more affordable alternative lipstick with the same shade as YSL 100. \\
    2. Lookalike-seeking queries: For example, the query ``lululemon lookalike yoga pants'' indicates the user's desire to find yoga pants from other brands that resemble Lululemon in style and fit. \\
    3. Question-answering queries: When a user enters ``what size for a 1.5-year-old boy'', the intent is to purchase clothing or footwear for a 1.5-year-old boy, requiring a recommendation based on appropriate age-specific sizing. \\
    4. General queries: For example, a query like ``dresses'' represents a standard product search, which can be processed using general retrieval and understanding methods. \\
    \\
    Based on the above examples, please analyze the following query: ``La Mer dupe'' \\
    \midrule
    \textbf{Response} \\
    The query ``La Mer dupe'' refers to products that deliver similar repair, hydration, and anti-aging benefits as La Mer, but at a lower price.
    La Mer is known for high-performance formulations, with flagship products like The Moisturizing Cream and The Concentrate positioned in the luxury segment. While its proprietary Miracle Broth™ remains unique, alternatives using peptides, ceramides, hyaluronic acid, centella asiatica, and niacinamide can offer comparable functional benefits. Below are recommended alternatives, categorized by product function. \\
    1. Alternatives to The Moisturizing Cream (Focus: Repair + Anti-Aging) \\
    Winona Barrier Repair Cream \\
    → Ideal for sensitive skin; strengthens the skin barrier with mild, effective ingredients. High value for daily repair. \\
    Proya Red Carpet Peptide Cream \\
    → Combines peptides and niacinamide for anti-aging. Rich yet non-greasy, priced at ~1/5 of La Mer. \\
    The Ordinary 10\% Niacinamide + 1\% Zinc (paired with moisturizer) \\
    → Budget-friendly option for oil control, brightening, and repair when layered. \\
    2. Alternatives to The Concentrate (Focus: Repair + Brightening) \\
    SkinCeuticals CE Ferulic Serum \\
    → Premium alternative with strong antioxidant and anti-photodamage effects; more affordable than La Mer. \\
    Runbaiyan Hyaluronic Acid Repair Serum \\
    → Simple, effective formula with hyaluronic acid and ceramides; suitable for daily barrier support. \\
    HBN Hyaluronic Acid Repair Serum \\
    → Contains hyaluronic acid and centella asiatica; targets overnight recovery and soothing, ideal for younger skin. \\
    Key Considerations \\
    La Mer's Miracle Broth™ has no direct equivalent. \\
    Functional similarity can be achieved through targeted ingredient combinations.
    Selection should be based on skin type (dry/oily/sensitive) and individual needs, not trends. \\
    \bottomrule
\end{tabular}
\label{tab:URP}
\end{table*}

\begin{table*}[h!]
\centering
\small
\caption{Detailed prompts for instructing Qwen3‑30B‑A3B‑Instruct in information extraction.}
\vspace{-10pt}
\begin{tabular}{p{0.9\linewidth}}
    \toprule
    \textbf{System Prompt} \\
    You are a data processing assistant specialized in the Chinese e-commerce domain, proficient in extracting keyword phrases from the reasoning outputs of LLMs that are closely relevant to users' search queries. \\
    \midrule
    \textbf{User Prompt} \\
    Please return the extraction results according to the following requirements: \\
    1. Output keyword phrases only—eliminate duplicates, avoid redundant prefixes, and exclude any terms unrelated to products or brands. \\
    2. If the user query explicitly targets a specific brand, do not include other brand names. \\
    3. For question-answering or dupe-seeking queries, extract the product/brand name or specific product attributes. \\
    4. If multiple keywords are extracted, separate them with ``,''. \\
    \\
    Now, given the following input, please extract the relevant keywords accordingly: \\
    Query: ``La Mer dupe'' \\
    Reasoning: ``LLMs reasoning output from the previous unconstrained reasoning phase'' \\
    \midrule
    \textbf{Response} \\
    Winona, Proya, The Ordinary, SkinCeuticals, Runbaiyan, Winona, HBN \\
    \bottomrule
\end{tabular}
\label{tab:IEP}
\end{table*}

\begin{table}[h!]
\centering
\small
\caption{Complete query‑side and item‑side inputs for the direct‑embedding dense retrieval model in constructing items set \ding{173}-\ding{172}.}
\vspace{-7pt}
\begin{tabular}{p{0.9\linewidth}}
    \toprule
    \textbf{Complete Query‑Side Input for Constructing Items Set \ding{172}} \\
    <|im\_start|>system\textbackslash n You are a helpful, respectful and honest assistant.<|im\_end|>
    <|im\_start|>user\textbackslash n La Mer dupe <|im\_end|>
    <|im\_start|>assistant\textbackslash n <|im\_end|> \\
    \midrule
    \textbf{Complete Query‑Side Input for Constructing Items Set \ding{173}} \\
    <|im\_start|>system\textbackslash n You are a helpful, respectful and honest assistant.<|im\_end|>
    <|im\_start|>user\textbackslash n La Mer dupe (Winona, Proya, The Ordinary, SkinCeuticals, Runbaiyan, HBN) <|im\_end|>
    <|im\_start|>assistant\textbackslash n <|im\_end|> \\
    \midrule
    \textbf{Complete Item‑Side Input} \\
    <|im\_start|>system\textbackslash n You are a helpful, respectful and honest assistant.<|im\_end|>
    <|im\_start|>user\textbackslash n Winona Soothing \& Hydrating Special Care Cream 5g + Repair Essence, Lightweight Sunscreen for Sensitive Skin <|im\_end|><|im\_start|>assistant\textbackslash n <|im\_end|> \\
    \bottomrule
\end{tabular}
\label{tab:deinput}
\end{table}

\begin{table}[h!]
\centering
\small
\caption{The relevance assessment process in TaoSR1.}
\vspace{-8pt}
\begin{tabular}{p{0.9\linewidth}}
    \toprule
    \textbf{Input} \\
    Human: <cot> \\
    Query:  La Mer dupe \\
    Item: La Mer essence lotion set with firming, moisturizing, soothing, and anti‑wrinkle skincare \\
    \midrule
    \textbf{Output} \\
    1. Query Intent Analysis \\
        Category Information: No explicit category indicated. \\
        Attributes: Brand (``La Mer''), Special Attribute (``dupe''/affordable alternative). \\
    2. Item Information Analysis \\
        Regarding the brand intent (``La Mer''), the item is explicitly described as ``La Mer.'' \\
        Regarding the ``dupe'' (affordable alternative) intent, the item description does not convey any indication of being a substitute or lower-cost alternative. Instead, it directly uses the original brand name, which contradicts the ``dupe'' intent. \\
    3. Category Matching \\
        The query expresses no explicit category intent. Therefore, category comparison is not applicable. (No comparison performed) \\
    4. Attribute Matching \\
        Brand: The query specifies ``La Mer'' and the item is labeled as ``La Mer'' → Fully Matched \\
        Special Attribute (Dupe): The query seeks an affordable alternative, but the item does not indicate cost-effectiveness or substitution. Moreover, the use of the official brand name in the title suggests it is an original product, which directly conflicts with the ``dupe'' requirement. → Complete Mismatch \\
    5. Final Relevance Assessment \\
        Rationale: Although the brand attribute is fully matched, the item fails to satisfy the ``dupe'' intent and explicitly promotes the original brand, which fundamentally contradicts the user’s request for an affordable alternative. According to the relevance tier mapping rules, when category matching is ``not applicable'' and attribute matching contains a ``complete mismatch'' the overall relevance level is classified as – Partial Mismatch. \\
        \\
        Relevance Level: – Partial Mismatch \\
    \bottomrule
\end{tabular}
\label{tab:reljudge}
\end{table}

\begin{table}[b!]
\centering
\small
\caption{Examples of constructed Query-CoT-Item triplets.}
\vspace{-10pt}
\begin{tabular}{p{0.9\linewidth}}
    \toprule
    \textbf{Query} \\
    La Mer dupe \\
    \midrule
    \textbf{CoT} \\
    Winona, Proya, The Ordinary, SkinCeuticals, Runbaiyan, HBN \\
    \midrule
    \textbf{Items} \\
    1. Winona Soothing \& Hydrating Special Care Cream 5g + Repair Essence, Lightweight Sunscreen for Sensitive Skin \\
    2. Winona Barrier Repair Cream 5g – Hydrating, Redness Relief \\
    3. Proya Ruby Peptide Cream 3.0 – Anti‑Wrinkle, Firming, Hydrating \& Plumping Skincare \\
    4. HBN Recovery Essence 2.0 – Pre‑Serum with Yeast, Hyaluronic Acid \& Ceramides for Soothing, Hydration \& Skin Barrier Repair \\  
    \bottomrule
\end{tabular}
\label{tab:triple_sample}
\end{table}

\begin{table}[b!]
\centering
\small
\caption{Complete query‑side and item‑side inputs for \modelname{} in the cold-start stage.}
\vspace{-10pt}
\begin{tabular}{p{0.9\linewidth}}
    \toprule
    \textbf{Complete Query‑Side Input for \modelname{}} \\
    <|im\_start|>system \\
    You are a query assistant for a Chinese e-commerce search platform, and you are adept at understanding various user queries.<|im\_end|>  \\
    <|im\_start|>user \\
    Please think and reason about the given query to fully understand the product search intent behind it. Consider and reason from multiple angles about various potential related phrases to the query. Do not include the query itself in the potential phrases you infer. Please output in the format: <think> Various Potential Related Phrases Associated with the Query <\textbackslash think><emb>. \\
    Query: La Mer dupe<|im\_end|> \\
    <|im\_start|>assistant \\
    <think> Winona, Proya, The Ordinary, SkinCeuticals, Runbaiyan, HBN <\textbackslash think><emb><|im\_end|>\\
    \midrule
    \textbf{Complete Item‑Side Input for \modelname{}} \\
    <|im\_start|>system \\
    You are a query assistant for a Chinese e-commerce search platform, and you are adept at understanding various products.<|im\_end|>  \\
    <|im\_start|>user \\
    Item: Winona Soothing \& Hydrating Special Care Cream 5g + Repair Essence, Lightweight Sunscreen for Sensitive Skin<|im\_end|> \\
    <|im\_start|>assistant \\
    <emb><|im\_end|>\\
    \bottomrule
\end{tabular}
\label{tab:coldstart_input}
\end{table}

\begin{table}[b!]
\centering
\small
\caption{Examples of CoTs sampled from \modelname{} during RL.}
\vspace{-10pt}
\begin{tabular}{p{0.9\linewidth}}
    \toprule
    \textbf{Query} \\
    Eliminate Climbing Ivy \\
    \midrule
    \textbf{Sampled CoTs} \\
    1. <think>Specialized Herbicide, Plant Inhibitor<\textbackslash think><emb> \\
    2. <think>Plant Growth Inhibitor, Weed Remover<\textbackslash think><emb> \\
    3. <think>Plant Pruning Shears, Eco-friendly Remover \\<\textbackslash think><emb> \\
    4. <think>Eco-friendly Vine Pruner, Herbicide<\textbackslash think><emb> \\
    5. <think>Weed Remover, Plant Restoration Activated Carbon<\textbackslash think><emb> \\
    6. <think>Herbicide, Biopesticide<\textbackslash think><emb> \\
    7. <think>Hexazinone, Hexazinone Powder, Plant Inhibitor \\<\textbackslash think><emb> \\
    8. <think>Weed Remover, Plant Repellent<\textbackslash think><emb> \\
    \bottomrule
\end{tabular}
\label{tab:cots_group}
\end{table}

\end{document}